\documentstyle[preprint,aps]{revtex}

\begin{document}

\draft

\title{Thermal divergences on the event horizons of two-dimensional
	black holes}

\author{Daniel J. Loranz\cite{Lor} and William A.\ Hiscock\cite{His}}

\address{Department of Physics, Montana State University, Bozeman, Montana
59717}

\author{Paul R.\ Anderson\cite{And}}

\address{Department of Physics, Wake Forest University, Winston-Salem, North
Carolina 27109}

\date{April 25, 1995}

\maketitle

\begin{abstract}

The expectation value of the stress-energy tensor $\langle
T_{\mu\nu}\rangle$ of
a free conformally invariant scalar field is computed in a general static
two-dimensional black hole spacetime when the field is in either a
zero temperature vacuum state or a thermal state at a nonzero temperature.
It is found that for every static two-dimensional black hole the stress-energy
diverges strongly on the event horizon unless the field is in a state at
the natural black hole temperature which is defined by the surface
gravity of the event horizon.  This implies that both extreme and nonextreme
two-dimensional black
holes can only be in equilibrium with radiation at the natural black hole
temperature.

\end{abstract}

\pacs{ }

\section{INTRODUCTION}
Since the discovery of black hole radiance two decades ago, the study of
quantum aspects of black holes has been regarded as one of the most
likely areas in which to gain further insights into the nature of quantum
gravitational processes. Recent work in several distinct areas has led to
a common central concern: does the stress-energy tensor of a quantized
field diverge on the event horizon? If so, is the divergence weak and
essentially ignorable, or is it strong, calling into question the very
existence of a meaningful semiclassical black hole solution?

It is well known that Schwarzschild and Reissner-Nordstr\" om black holes have
a well defined temperature which is related to their surface gravity.  If
quantized fields are in a thermal state at this temperature (this is the
Hartle-Hawking state), then one expects that the stress-energy is finite
on the past and future event horizons \footnote {In this paper, we shall be
concerned only with possible equilibrium states of the black hole-quantized
field system. We shall not consider states such as the Unruh vacuum state,
which is appropriate to a black hole formed by collapse, and which would be
regular on the future event horizon but would diverge on the past event
horizon}.
This has been confirmed
by numerical calculations in four dimensions\cite{How,AHS}.  If the fields
are not in the Hartle-Hawking state, even if they are in a thermal state
at some other temperature, then the stress-energy diverges severely on the
event horizon of the black hole. This strong divergence indicates that no
solution to the quantum or semiclassical theory would be ``near'' the
classical solution in these cases; the backreaction to the
quantized fields would profoundly affect the geometry in a nonperturbative
fashion if the fields are not in the Hartle-Hawking state.

Despite the classic black hole uniqueness theorems, there are several
reasons to be interested in quantum effects in more general black hole
spacetimes.
For example backreaction effects due to the nonzero stress-energy of the
quantum fields will alter the spacetime geometry near the event horizon of a
Schwarzschild or Reissner-Nordstr\" om black hole \cite{York,HKY,AHWY}.
Thus self-consistent solutions to the semiclassical backreaction equations
will not be described by the exact classical geometries. Inclusion of
additional fields, such as the dilaton suggested by superstring theories, may
also necessitate examining a larger class of black hole metrics. Within the
larger, general class of black hole metrics, those metrics which represent
extreme black holes (usually defined as having a degenerate horizon
structure and zero surface gravity) are of particular interest due to their
possible stability against evaporation.

Extreme black holes play a very important
role in certain contemporary investigations.  One such area is the
study of information loss due to the evaporation process \cite{HarStr}.
By studying the absorption and re-emission of radiation by an
initially extreme black hole, the issues of Planck scale physics may be
avoided, while capturing the essence of information loss. The simplicity
of semiclassical theories in two dimensions allows for the explicit solution
of such models\cite{CGHS}.

A second area involves the investigation of pair creation of magnetically
charged Reissner-Nordstr\" om black holes by an external magnetic field
\cite{Gi,GaSt,DGGH}. A curious discrepancy has been found between the pair
creation rate for extreme and
nonextreme black holes. This discrepancy can be understood simply if extreme
Reissner-Nordstr\" om black holes are assigned zero entropy (notwithstanding
their nonzero horizon area). Hawking, Horowitz, and Ross \cite{HHR} have
recently pointed out that in the Euclidean sector, the distance to the
horizon of an extreme black hole is infinite in all directions (as opposed to
merely in spacelike directions in the Lorentzian metric); the Euclidean
geometry may then be identified with an arbitrary period without the penalty
of introducing a conical singularity at the horizon. If one can indeed
assign an extreme black hole an arbitrary temperature, it follows that the
entropy will formally be zero, explaining the pair creation rate discrepancy.

Extreme dilaton black holes \cite{GiMa,GaHo} may play an important role
in superstring theories as representations of massive single string states
\cite{HolWil,DuRa}. These black holes are extreme in the sense that any
increase in the charge of the hole would result in a nakedly singular
spacetime. However, unlike the ordinary Kerr-Newman extreme black hole
metrics of general relativity, these black holes may have zero, finite,
or infinite surface gravities (and hence temperatures),
depending on the value of $a$, the dilaton coupling.
However, the assignment of thermodynamic properties such as temperature
and entropy to elementary particles (strings) is problematic, particularly
when infinite temperatures are contemplated. In order to avoid confusion,
we shall hereafter reserve the use of the adjective "extreme" for those
black holes (with or without dilaton fields) which have zero surface gravity,
unless otherwise explicitly stated.

Finally, Trivedi \cite{Triv} has shown that the stress-energy of
a quantized conformally coupled massless field has a weak divergence
on the event horizon of (almost) any extreme two-dimensional
black hole if the field is in a zero temperature vacuum state.   We have
recently shown \cite{AnHi} that this divergence does not occur for the four-
dimensional extreme Reissner-Nordstr\" om black hole
and that no divergence in the stress-energy occurs on the horizon
if the field is in a zero temperature vacuum state.  However,
if the field is in a thermal state at any nonzero temperature a severe
divergence in the stress-energy does occur on the horizon.
This means that there is a well defined temperature
(zero) for the extreme Reissner-Nordstr\" om black hole and it cannot
be coupled to radiation at an arbitrary temperature as suggested by
Hawking, Horowitz and Ross \cite{HHR}.

These studies raise several related natural questions concerning the
stress-energy of quantized fields in black hole spacetimes. Does the
divergence discovered by Trivedi exist in four-dimensional extreme black
hole spacetimes other than the extreme Reissner-Nordstr\" om spacetime?
Is the stress-energy of quantized fields in nonextreme black hole spacetimes
always finite on the event horizon if the fields are in the Hartle-Hawking
state and does it always diverge otherwise? Can other extreme black holes
be coupled to radiation at an arbitrary temperature even if the extreme
Reissner-Nordstr\" om black hole cannot?

The answers to these questions will have important implications.  For
example if the divergence found by Trivedi for extreme black holes existed for
some four-dimensional extreme black holes then quantum effects would greatly
alter the spacetime geometry near the horizons of such black holes, even though
all curvatures in that region may be far smaller than the Planck scale.
On the other hand, if one could assign an extreme black hole a temperature
other than that defined by the surface gravity, this would possibly allow one
to avoid infinite temperatures in dilatonic extreme black holes, and would also
allow the assignment of zero entropy to an extreme black hole with nonzero
horizon surface area, as proposed by
Hawking, Horowitz, and Ross.

The recent development of a method of numerically computing the stress-energy
of quantized scalar fields in an arbitrary static spherically symmetric
four dimensional spacetime \cite{AHS} has made it possible to address
these questions directly in four dimensions; as noted above they have
already been answered for Schwarzschild, Reissner-Nordstr\" om, and  extreme
Reissner-Nordstr\" om black holes.
However, the numerical methods do not allow one to examine arbitrary
black holes in four dimensions.  Instead one must numerically compute the
stress-energy for one black hole geometry at a time.

For this reason it is of interest to look at the two dimensional case.
Here the stress-energy can be computed analytically for a conformally
invariant field in an arbitrary two dimensional spacetime.  Thus one
can examine all static two dimensional black hole geometries.  In this paper
we compute the stress-energy for a conformally invariant
field in a general two dimensional static black hole spacetime.  We then
investigate under what conditions the stress-energy is finite or, at most,
weakly divergent on the event horizon.  We find in all cases
that the stress-energy tensor diverges strongly on the horizon unless the
fields are in a thermal state with a temperature equal to the natural
temperature defined by the surface gravity of the black hole.
Hence, the divergence we previously found in the four-dimensional extreme
Reissner-Nordstr\" om spacetime is of precisely the same form as that which
occurs when any two-dimensional black hole, extreme or nonextreme, is assigned
an ``unnatural''  temperature.

For two-dimensional extreme dilaton black holes, this implies that one
cannot escape the divergent behavior associated with infinite temperatures
by choosing to identify the Euclidean metric with a different period (and
hence temperature). Any such attempt will result in a strongly divergent
stress-energy tensor on the black hole's horizon.

Our sign conventions and notation follow Misner, Thorne and Wheeler
\cite{MiTh}; we also use natural units ($G = c = \hbar = k_B = 1$) throughout.

\section{CALCULATION OF $\left\langle T_{\mu\nu}\right\rangle$}
The most general two dimensional static spacetime metric may be written
in the form
\begin{equation}
	ds^2 = -F(R)dt^2 + {1 \over H(R)}dR^2 .
	\label{Eq1}
\end{equation}
It is always possible to transform this metric to ``Schwarzschild gauge'' by
defining a new spatial coordinate
\begin{equation}
	r = \int {\left( {F \over H} \right)}^{1/2} dR,
	\label{Eq2}
\end{equation}
which yields as the general metric form
\begin{equation}
	 ds^2 = -f(r)dt^2 + {1 \over f(r)}dr^2,
	  \label{Eq3}
\end{equation}
where $f(r)$ is an arbitrary function of $r$. This metric possesses
horizons at the locations $r_n$, where $f(r_n) = 0$, $(n=0,1,2,...)$.
Markovic and Poisson \cite{MaPo} have recently discussed stress-energy
divergences on Cauchy horizons in two-dimensional spacetimes.
In this paper we consider black holes with an arbitrary number
of horizons, but we restrict our attention to the outer
event horizon at $r = r_0$.

The conservation of stress-energy in the metric of Eq.(3) yields the
differential equations
\begin{equation}
	{T_{t}^{r}}_{,r} = 0,
	\label{Eq4}
\end{equation}
and
\begin{equation}
	{T_{r}^{r}}_{,r} + {1 \over 2}{f' \over f}T_{r}^{r}
	- {1 \over 2}{f' \over f}T_{t}^{t} = 0,
	\label{Eq5}
\end{equation}
where a prime denotes differentiation with respect to $r$.
Hereafter we will generally suppress the expectation value brackets for
simplicity of notation. Eq.(5) may be simplified by rewriting $T_{t}^{t}$
in terms of the trace
of the stress-energy tensor: $T_{t}^{t}= T_{\alpha}^{\alpha} - T_{r}^{r}$.
The conservation equations are then easily integrated;
\begin{equation}
	 T_{t}^{r} = C_1,
	\label{Eq6}
\end{equation}
and
\begin{equation}
	 T_{r}^{r} = {C_2 \over f} +
	{1 \over 2f}\int_{r_0}^{r} f'T_{\alpha}^{\alpha}dr.
	\label{Eq7}
\end{equation}
Eqs.(6,7) are the complete solution to the conservation of stress-energy
equations in two dimensions. The components of the stress-energy tensor in
a given spacetime then depend on one function, $T_{\alpha}^{\alpha}$, and
two integration constants, $C_1$ and $C_2$.

If the field is chosen to be conformally invariant, then the trace is given
by the conformal anomaly, $T_{\alpha}^{\alpha} = {R \over {24\pi}}$,
where $R$ is the Ricci scalar, which becomes
\begin{equation}
	T_{\alpha}^{\alpha} = -{f'' \over 24{\pi}}
	\label{Eq8}
\end{equation}
for the metric of Eq.(3). In the case of a conformally invariant field, all
information concerning the quantum state of the field is then encoded in the
two integration constants. Substituting the trace anomaly from Eq.(8) into
Eq.(7), one can explicitly perform the integral to find
\begin{equation}
	T_{r}^{r} =
	{C_2 \over f}-{f'^2 \over 96{\pi}f}+
	{{\pi} \over 6f}\left(T_{Hawking}\right)^2,
	\label{Eq9}
\end{equation}
where we have defined $T_{Hawking}$ in terms of the surface gravity, $\kappa$,
of the horizon at $r_{0}$,
\begin{equation}
	T_{Hawking}= {\kappa \over {2\pi}}
	= {f'|_{r_0} \over {4\pi}}.
	\label{Eq10}
\end{equation}
Note we have not assumed that $T_{Hawking}$ represents the physical
temperature of the black hole; rather we have simply used a familiar
definition to simplify a collection of terms involving $f'\left({r_0}\right)$.

The integration constants are fixed by choosing a particular quantum
state for the field. We will consider a state in which
the black hole is in thermal equilibrium with a surrounding heat bath.
The requirement of thermal equilibrium implies that $T_{\mu}^{\nu}$ must
be invariant under time reversal, and thus $T_{t}^{r} = C_1 = 0$.

The remaining integration constant, $C_2$, is now determined by fixing the
form of $T_{r}^{r}$ in an asymptotically flat region far from the horizon,
where $f \to constant$. (The apparently more general asymptotically
flat form $f' \to constant$ simply amounts to choosing asymptotically Rindler
coordinates rather than Minkowski.) We assume that the stress-energy
approaches the form appropriate to a two-dimensional gas of massless scalar
bosons at temperature $T$ far from the horizon,
\begin{equation}
	T_{r}^{r} \to {{\pi} \over 6} { T }^2,
	\label{Eq11}
\end{equation}
as the metric becomes asymptotically flat. Evaluating Eq.(9) in the
asymptotically flat region and using Eq.(11), we find
\begin{equation}
	C_{2} =
	{{\pi} \over 6} T^{2}-
	{{\pi} \over 6} \left(T_{Hawking}\right)^2.
	\label{Eq12}
\end{equation}

\section{$\langle T_{\mu\nu} \rangle$ ON THE HORIZON}

Having completely integrated the conservation equation and solved for the
stress
-energy tensor, we turn to the issue of its regularity on the event horizon.
Since the coordinate system used in the metric of Eq.(3) is singular on the
event horizon at $r_0$, we will evaluate the stress-energy tensor components
in an orthonormal frame attached to a freely falling observer.
The basis vectors of the frame are chosen to be the two-velocity $
e_0^\alpha = u^{\alpha}$ and a unit length spacelike vector
$e_1^\alpha = n^{\alpha}$ orthogonal to $u^{\alpha}$, so
that $n^{\alpha}u_{\alpha} = 0$ and $n^{\alpha}n_{\alpha} = +1$.
Using the timelike Killing vector field to define a conserved energy, the
geodesic equation may then be solved to find
\begin{equation}
	u^{t} = {\gamma} / f ,\quad u^{r} = -\sqrt{{\gamma}^2-f}
	\label{Eq13}
\end{equation}
and
\begin{equation}
	 n^{t} = - {1 \over f} \sqrt{{\gamma}^2-f}
	,\quad n^{r} = {\gamma}.
	\label{Eq14}
\end{equation}
where $\gamma$ is the energy per unit mass along the geodesic.
The components of the stress-energy tensor in the freely-falling orthonormal
frame are then given in terms of the coordinate components by
\begin{equation}
	T_{00} = {\gamma^2 \left(T_{r}^{r}-T_{t}^{t} \right) \over f}-
	T_{r}^{r},
	\label{Eq15}
\end{equation}
\begin{equation}
	T_{11} = {\gamma^2 \left(T_{r}^{r}-T_{t}^{t} \right) \over f}
	+T_{t}^{t},
	\label{Eq16}
\end{equation}
and
\begin{equation}
	T_{01} = - {\gamma\sqrt{\gamma^2-f} \left(T_{r}^{r}-T_{t}^{t}
	 \right) \over f}.
	\label{Eq17}
\end{equation}
Since the value of $\gamma$
is arbitrary, the stress-energy will be regular on the horizon only if
$T_{t}^{t}$, $T_{r}^{r}$, and the combination $\left(T_{r}^{r}-T_{t}^{t}
\right)/f$ are each separately finite at $r_0$.  Because a possible divergence
in either $T_{t}^{t}$ or $T_{r}^{r}$ will be made stronger by the extra
$f^{-1}$
in the combination $\left(T_{r}^{r}-T_{t}^{t} \right)/f$, we will focus on this
combination as representing the strongest possible divergence in $\langle
T_{\mu
\nu}\rangle$. Using Eqs.(9,10) and the trace anomaly, this combination of
terms may be written as
\begin{equation}
	{ T_{r}^{r}-T_{t}^{t} \over f} =
	{2C_2 \over f^2}-{1 \over 48{\pi}}{{f'^2-f'^2}|_{r_0}
	-2ff'' \over f^2}.
	\label{Eq18}
\end{equation}
The second term on the right hand side of Eq.(18) is
0/0 on the event horizon.  Applying l'Hospital's rule to this term and
rewriting
$\left(T_{r}^{r}-T_{t}^{t} \right)/f$ in the limit as $r \to r_0$ gives
\begin{equation}
	\lim_{r \to r_0} {{T_{r}^{r}-T_{t}^{t}} \over {f}} =
	\left({{2C_2 \over f^2}+
	{1 \over 48{\pi}}{f''' \over f'}}\right)\Bigg|_{r_0}.
	\label{Eq19}
\end{equation}
For an extreme black hole (by the conventional definition, i.e., one
with zero surface gravity), $f'|_{r_0} = 0$, and the second
term of Eq.(19) diverges. This is the unavoidable divergence Trivedi previously
discovered \cite{Triv}. The only escape from this divergence would be if $f'''$
approaches zero as fast or faster than $f'$ does in the limit $r \rightarrow
r_0$.\footnote {If one examines a more general, non-conformally coupled field
(so that the trace of the stress-energy tensor is not given by Eq.(8)), then
the
condition for regularity of the stress-energy on the event horizon is that
the limit of ${\left({T_\alpha}^\alpha \right)'}/ f'$ as
$r \rightarrow r_0$ must be finite.} Of course two-dimensional extreme
black hole metrics for which this occurs form a set of measure
zero in the space of all extreme two-dimensional black hole metrics.
However, it remains to be seen whether two-dimensional gravitational
dynamics, when semiclassical backreaction is included,  might cause extreme
black hole solutions to evolve towards such a state.

If $C_2 \neq 0$, then there is a far more serious divergence of the stress-
energy tensor on the event horizon. The energy density and pressure seen by
an infalling observer will diverge as $f^{-2}$, irrespective of whether the
black hole is extreme or not. However, we have previously seen that the
integration constant $C_2$ may be expressed in terms of the difference
between the square of the ``natural'' temperature of the black hole,
$T_{Hawking}$, defined by the surface gravity,
and the square of the asymptotic temperature assigned to the black hole, $T$,
as shown in Eq.(12). We thus see that unless the temperature assigned to
the black hole, $T$, is precisely equal to the natural temperature defined
by the surface gravity, $T_{Hawking}$, the stress-energy of a quantized field
will diverge strongly on the event horizon. Further, the form of the divergence
is independent of whether the black hole is extreme or nonextreme; extreme
(zero surface gravity) black holes have a natural temperature, namely, zero,
in precisely the same fashion as nonextreme
holes, and may not be assigned arbitrary temperature without serious
consequences. The divergence of $\langle T_{\mu\nu} \rangle$ on the
event horizon of an extreme Reissner-Nordstr\" om black hole in
four dimensions when the temperature is chosen to be other than zero
\cite{AnHi} is thus seen to be simply an example of the general strong
divergence which occurs for all black holes when assigned an
inappropriate temperature.

In conclusion, the stress-energy tensor of a conformally coupled quantized
field will diverge strongly on the event horizon of any two-dimensional black
hole unless the temperature of the black hole is chosen to be equal to the
natural temperature defined by the surface gravity of the horizon. If the
temperature is chosen to be the natural value, then the stress-energy tensor
will be regular on the horizon unless the black hole is extreme. If the black
hole is extreme, then there will be a weak  divergence of the
stress-energy on the horizon, except in a set of metrics of measure zero.
Whether two-dimensional extreme black hole metrics evolve naturally towards
these non-divergent cases when semiclassical backreaction is included remains
to be determined.

\acknowledgements
The work of W.\ A.\ H. was supported in part by National Science Foundation
Grant No.  PHY92-07903

\end{document}